\documentstyle[12pt,aasms4]{article}

\received{16 November 1995}
\accepted{24 June 1997}

\begin{document}

\title{Quasar-Galaxy Associations from Gravitational Lensing: Revisited}

\author{Zong-Hong Zhu}
\affil{Department of Astronomy, Beijing Normal University, 
       Beijing 100875, China}

\author{Xiang-Ping Wu}
\affil{Department of Physics, University of Arizona, Tucson, AZ 85721 and \\
       Beijing Astronomical Observatory, Chinese Academy of Sciences,
       Beijing 100080, China}

\and

\author{Li-Zhi Fang}
\affil{Department of Physics, University of Arizona, Tucson, AZ 85721}

\begin{abstract}
The theoretically expected amplitude of 
the associations of background quasars with foreground galaxies
as a result of gravitational lensing has been updated in this paper.  
Since the galactic matter alone yields an amplitude of 
quasar overdensity smaller than that observed, 
a special attention has been paid to the examination or re-examination of 
the uncertainties in the estimate of the quasar enhancement factor arising
from the cosmic evolution of galaxies, the core radius and velocity
bias of galactic matter distributions,   
the clusters of galaxies, the obstruction effect by
galactic disks, the non-zero cosmological constant, etc. 
Unfortunately, none of these factors has been shown to be able to  
significantly improve the situation, although
a combination of some effects may provide a result that 
marginally agrees with observations. 
It is concluded that the quasar-galaxy association  
still remains to be an unsolved puzzle in today's astronomy, 
if the reported quasar-galaxy associations are not due to the statistical 
variations and/or the observed 
quasar number counts as a whole have not been seriously contaminated by 
gravitational lensing.
\end{abstract}


\section{Introduction}

While there are increasing observational evidences for  
the associations between background quasars and foreground objects 
such as quasars, galaxies, groups and
clusters of galaxies, the theoretical explanations still remain 
unsatisfactory (see Wu 1996 for a recent review). This has led to
the longstanding argument for a non-cosmological origin of quasar redshifts. 
Nowadays, it is most likely and also widely
accepted that the overdensity of high-redshift quasars behind  
low-redshift objects is relevant to the magnification effect by 
the gravitational lensing of the foreground objects, 
though theoretical studies have always found a relatively weak amplitude 
as compared with observations. In this paper, we intend to update 
the theoretical estimate of the amplitude of the quasar-galaxy 
associations in the framework of gravitational lensing, 
taking into account the influence of various factors such as
the galactic matter distributions (e.g. core radius and 
velocity bias parameter), the galactic morphologies, 
the galaxy evolution with cosmic epoch, the environmental matter 
contributions of galaxies from their host clusters,  the non-zero cosmological 
constant, etc. Through the present work we would like  
to demonstrate how our predictions of the quasar-galaxy associations
are affected by these factors. Eventually, 
we would re-examine the question whether the quasar-galaxy associations can be 
interpretated as the result of gravitational lensing.

The amplitude of the quasar-galaxy associations is often characterized 
by the so-called enhancement factor $q$, which is the ratio of the
disturbed or observed surface number density of quasars (galaxies)
to the undisturbed or intrinsic value over a given area around 
galaxies (quasars).
Note that the quasar enhancement factor $q_q$ is often used in theoretical
studies while observations actually provide the galaxy enhancement
factor $q_g$. We make no distinction below between these two parameters. 
In 1989 Narayan introduced a simple and elegant formula 
in the scenario of gravitational lensing to estimate $q$:
\begin{equation}
q=\frac{N_q(<B+2.5\log \mu)}{N_q(<B)}\;\frac{1}{\mu},
\end{equation}
where $N_q$ is the cumulative quasar number count above the limiting magnitude
$B$,  $\mu$ is the magnification factor and  $2.5\log\mu$ and $1/\mu$ 
account for the magnification bias and the area distortion 
because of the gravitational deflection of light, respectively. 
The advantage of this method is that it is independent of
specific lensing models and hence,  applicable to 
various matter distributions. Moreover, it avoids the introduction
of the lensing magnification probability and the employment of
the quasar luminosity function and therefore, simplifies considerably 
the theoretical computations.  However, one should be cautious 
of applying this expression for the actual observations: Eq.(1)
is valid only for a single galaxy and a 
given magnification which generally acts as a function of the
searching distance from the galaxy. 
While the measurement of the quasar-galaxy
associations is made over a certain area around
an ensemble of galaxies or quasars,   
a statistically expected enhancement factor $\langle q\rangle$ 
needs to be found in order to compare with observations.

We summarize in Table 1 the updated observations of the quasar-galaxy 
associations and their resulted enhancement factors.
In the present paper,  we only concentrate on the optically-selected 
quasars. It is immediately apparent
from Table 1 that observations provide both positive
and negative results for the quasar-galaxy associations. In a sense,
this is probably representative of the signature of 
gravitational lensing (Wu 1994). 
Indeed, previous theoretical studies of the phenomenon
in terms of gravitational lensing 
could give rise to a scenario that is essentially consistent with the major
features of the observations if minor modifications to the
conventional lensing models were made. 
For instance, one may achieve the observed enhancement factors
by requiring rather a large galaxy velocity dispersion 
(Webster et al. 1988; Narayan 1989)
or rather a steep intrinsic quasar luminosity function 
(Bartelmann \& Schneider 1993). Yet, there are good reasons to
believe that gravitational 
lensing should be the natural cause for the quasar-galaxy 
associations, and the inefficiency of the current lensing
explanations may arise from our poor understanding of the 
various effects in modeling of galaxies as lenses.

\placetable{table-1}

We describe the formulas for the estimate of 
the expected quasar enhancement factor in section 2. 
In section 3 we investigate the contributions of clusters
of galaxies. The numerical computations are carried out in section 4 and 
our main results are summarized in section 5.
Throughout the paper, we adopt a Hubble constant of 
$H_0=50$ km s$^{-1}$ Mpc$^{-1}$ and a flat cosmological
model of $\Omega_0+\lambda_0=1$, where $\Omega_0$ and 
$\lambda_0$ denote the matter and cosmological constant
contributions, respectively.

\section{Galaxies as lenses}

It is customary to assume that 
the contamination of gravitational lensing in the quasar number counts
as a whole is negligible. This enables us to use the observed quasar
number-magnitude relation $N_q(<B)$ to represent approximately the   
intrinsic quasar number count in Eq.(1). Numerous quasar surveys 
using ultraviolet excess and slitless spectroscopy basically yield
the similar surface number density of quasars for a given limiting 
magnitude (V\'eron 1993; references therein). 
Our best-fit quasar number-magnitude relation reads
\begin{equation}
\begin{array}{ll}
N_q(<B)=\alpha_1\beta_1 10^{(B-19.14)/\beta_1},  & 
                    (0.2<z_s<2.2,\;B<19.14);\\
N_q(<B)=\alpha_1\beta_1+\alpha_2\beta_2(10^{(B-19.14)/\beta_2}-1),  & 
                    (0.2<z_s<2.2,\;B>19.14);\\
N_q(<B)=\alpha_3\beta_3 10^{(B-19.14)/\beta_3},  &
                    (2.2<z_s<3.0,\;B<22),\\ 
\end{array}
\end{equation}
where $(\log\alpha_1,\beta_1^{-1})=(0.62\pm0.07,0.95\pm0.04)$, 
$(\log\alpha_2,\beta_2^{-1})=(0.61\mp0.04,0.21\pm0.03)$, 
$(\log\alpha_3,\beta_3^{-1})=(-0.46\mp0.02,0.70\pm0.05)$. 
Since the evaluation of the quasar enhancement factor $q$ from Eq.(1) 
depends critically on the adopted $N_q(<B)$, we have included 
the uncertainties in the fit of $N_q(<B)$.  Alternatively,
we have  excluded the variability-selected quasars in order
to keep the same selection criteria of quasar samples as those 
used in the measurements of the quasar-galaxy associations (Table 1).
Recall that the variability-selected quasars
show a  number-magnitude relation without a turnover around 
$B\approx19.14$ (Hawkins \& V\'eron 1993), resulting in a relatively
small value of $q$  (Wu 1994). Also, 
it is necessary to note that Eq.(2) is valid within $B<22$.

We use two types of density profiles to model galaxy matter distribution:
a singular isothermal sphere (SIS) and a softened SIS with a core 
radius $r_c$ (ISC): 
\begin{equation}
\begin{array}{ll}
\rho=\frac{\sigma_{v}^2}{2\pi G}\frac{1}{r^2}, & {\rm SIS};\\
\rho=\frac{\sigma_{v}^2}{2\pi G}\frac{1}{r^2+r_c^2}, & {\rm ISC},
\end{array}
\end{equation}
in which $\sigma_{DM}$ is the one-dimensional velocity 
dispersion of galactic dark matter. Their resulting 
lensing magnifications are simply
\begin{equation}
\begin{array}{ll}
\mu=\frac{\theta}{\theta-\theta_E}, & {\rm SIS};\\
\mu=\left|\left(1-\theta_E\frac{\sqrt{\theta^2+\theta_c^2}-\theta_c}
			{\theta^2}\right)
          \left(1+\theta_E\frac{\sqrt{\theta^2+\theta_c^2}-\theta_c}
		        {\theta^2}-\theta_E\frac{1}
                                   {\sqrt{\theta^2+\theta_c^2}}
          \right)
    \right|^{-1}, & {\rm ISC},
\end{array}
\end{equation} 
where $\theta_E=4\pi(\sigma_v/c)^2(D_{ds}/D_s)$ is the Einstein radius
by SIS, $\theta_c\equiv r_c/D_d$,  and $D_d$, $D_s$ and $D_{ds}$
are the angular diameter distances to the galaxy  at redshift $z_d$, 
to the background quasar at redshift $z_s$, 
and from the galaxy to the quasar, respectively.

To compare with observations, we use the integral form of Eq.(1), i.e., 
we estimate the average quasar enhancement factor over an area  
from $\theta_1$ to $\theta_2$ around a foreground galaxy: 
\begin{equation}
\overline{q}(B,\theta_1,\theta_2)=\frac{
2\int_{\theta_1}^{\theta_2}\;q_{\rm local}(B,\theta)\;\theta d\theta}
{\theta_2^2-\theta_1^2},
\end{equation}
where $q_{\rm local}$ denotes the ``local'' value given by Eq.(1).
For  an ensemble of galaxies as lenses we adopt the Schechter function
at $z\approx0$: 
$\phi(L,0)dL=\phi_*(L/L_*)^{\alpha}\exp(-L/L_*)dL/L_*$.
This expression can be converted into the velocity dispersion 
distribution through the empirical formula between galactic luminosity 
and central velocity dispersion, namely,
the Faber-Jackson relation for early-type galaxies (E/S0)
$L/L_*=(\sigma_v/\sigma_*)^4$ and the Tully-Fisher relation for spiral 
galaxies (S) $L/L_*=(\sigma_v/\sigma_*)^{2.6}$, where
$\sigma_*$ is the characteristic velocity dispersion 
corresponding to an $L_*$ galaxy. The total population of galaxies
with $L>L_{min}$ around redshift $z_d$ is thus
\begin{equation}
\langle n_g\rangle=
                   \displaystyle\sum_{i}\int_{L_{min,i}}^{\infty}
	           \gamma_i\phi_i(L,z_d)dL,
\end{equation}
in which $i$ and $\gamma_i$ represent, respectively, the $i$-th
morphological type and composition of galaxies. In the following
computation, we adopt the parameters ($\sigma_*$, $L_*$, $\alpha$) 
given by Efstathiou, Ellis, \& Peterson (1988) in $B_T$ band 
(see also Fukugita \& Turner 1991) and the morphological 
composition E:S0:S=12:19:69 found by Postman \& Geller (1984).

We now deal with the spatial distribution of galaxies.
The proper distance within redshift $dz_d$ of $z_d$ 
in an $\Omega_0+\lambda_0=1$ universe is given by
\begin{equation}
dr_{prop,z_d}=\frac{c}{H_0}\frac{dz_d}
{(1+z_d)\sqrt{\Omega_0(1+z_d)^3+1-\Omega_0}},
\end{equation}
and the angular diameter distance from $z_1$ to $z_2$ is thus
\begin{equation}
d(z_1,z_2)=\frac{c}{H_0}\frac{1}{1+z_2}\int_{z_1}^{z_2}
\frac{dz}{\sqrt{\Omega_0(1+z)^3+1-\Omega_0}}
\end{equation}
so that the angular diameter distances to the foreground galaxy ($D_d$), 
to the background quasar ($D_s$) and from the galaxy to the quasar ($D_{ds}$)
are $D_d=d(0,z_d)$, $D_s=d(0,z_s)$ and $D_{ds}=d(z_d,z_s)$, respectively.

Because the galaxies in the measurements of quasar-galaxy 
associations are most likely located at redshifts between $\sim0.1$ 
and $\sim1$, we need to include the effect of galaxy evolution. 
For simplicity, we consider two types of empirical evolutionary 
models: 
(I) the luminosity-dependent evolution by Broadhurst et al. (1988) and 
(II) the galaxy merging model by 
Broadhurst et al. (1992). The increase of galaxy number density 
$\phi_i(L,z_d)$ with redshift $z_d$ in Model I  is given by
\begin{equation}
\log\phi_i(L,z_d)=\log\phi_i(L,0)+(0.1z_d+0.2z_d^2)
                  \log\frac{\phi_i(L,0)}{\phi_i(L_{max},0)},
\end{equation}
in which $L_{max}$ is the truncated luminosity at $M_{max}=-23.5$, while 
for Model II, 
\begin{equation}
\begin{array}{l}
\phi_i(L,z_d)=f(z_d)\phi_i(L,0),\\
f(z_d)=\exp\{-Q[(1+z_d)^{-\beta}-1]/\beta\},
\end{array}
\end{equation}
where $Q$ describes the galaxy merging rate and $\beta$ is the ratio of
the Hubble constant to the age of the universe. Additionally, the galaxy
velocity dispersion in Model II varies with $z_d$ as 
$\sigma(z_d)=\sigma(0)f(z_d)^{-\nu}$. In the following calculations
we choose $Q=4$, $\beta=3/2$ and $\nu=1/4$ (see also Rix et al. 1994).
As a comparison, the velocity dispersion of galaxies in Model I
remains unchanged with redshift. This would result in a large population of
massive galaxies at high redshifts 
when coupled with the local Faber-Jackson and 
the Tully-Fisher relations. So, the effect of gravitational lensing 
by galaxies predicted in Model I may be overestimated. It should be also 
noted that both of the evolutionary models employed here are inapplicable
to the universe with a nonzero cosmological constant. In general, 
the introduction of $\lambda_0$ would increase the comoving volume and
hence reduce the galaxy number density. On the other hand, $\lambda_0$
would increase the estimate of the intrinsic luminosity of galaxies.
Here, we do not intend to explore an evolutionary model of galaxies
for a $\lambda_0\neq0$ universe. Instead, we just give a caution that 
the influence of the galaxy evolution described by Model I and II
and the nonzero cosmological constant upon the quasar-galaxy associations 
can not be simultaneously taken into account.

Noticing that the association areas are very close to
the central regions of foreground galaxies in some cases (Table 1),  
we give a ``maximum'' estimate of the possible effect of  the ``obstruction''
by the luminous disks of galaxies on the selection of background
quasars. To this end, we assume a face-on circular disk with radius $R_d$ 
for the luminous area of a galaxy. We then require that 
the search distance around a galaxy should be larger than $R_d$ in order
to detect the background quasars.  
For a uniform surface brightness disk with luminosity $L$, 
we utilize $R_d=14 (L/L_*)^{1/2}$ kpc to compute the corresponding 
obstructing size of a galaxy (Grossman \& Narayan 1988).

Finally, the expected enhancement factor of the quasar surface number 
density around foreground galaxies is obtained by 
\begin{equation}
\langle q\rangle(B,z_s,\theta_1,\theta_2,m_g)=
\frac{\int_0^{z_s}\;4\pi D_d^2(1+z_d)^3\;
\langle qn_g\rangle\;dr_{prop,z_d}}
{\int_0^{z_s}\;4\pi D_d^2(1+z_d)^3\;
\langle n_g\rangle\;dr_{prop,z_d}},
\end{equation}
where
\begin{equation}
\langle qn_g\rangle=\displaystyle\sum_{i}\int_{L_{min,i}}^{\infty}
	\overline{q}\gamma_i\phi_i(L,z_d)dL,
\end{equation}
and $m_g$ denotes the galaxy limiting magnitude used in 
the searches for
the quasar-galaxy associations and is related to the low
integration limit $L_{min}$ in Eq.(6) via 
\begin{equation}
m_g=M_*-2.5\log\frac{L_{min}}{L_*}+K(z_d)
+5\log\frac{(1+z_d)^2D_d}{10{\rm pc}}.
\end{equation}
Since the selections of galaxies are often made in the $R$ band, we 
employ the $K$ correction by Coleman, Wu, \& Weedman (1980).
Alternatively,
we choose intrinsic colors $B_T-R=(1.51,0.83)$ for (E/S0,S) to 
transform the luminosity function from $B_T$ band into
$R$ band: $M_*=-19.9_{-0.2}^{+0.4}-1.5+(R-B_T)$. 
For background quasars, an approximate color transformation of
$B-V\approx0.4$ is taken as an average value.

\section{Cluster contributions}

Galaxies are not isolated objects in the universe. We classify
galaxies as the cluster populations that trace the gravitational
potential of their host clusters and the field populations that
follow the large-scale structures of the universe.  
The contribution of the galaxy environmental matter from 
large-scale structures to the computation of the quasar enhancement 
factor has been shown to be negligible (Wu et al. 1996; 1997). Here 
we focus the effect of cluster matter on $\langle q\rangle$.

For the galaxies bounded in the gravitational potential of their
host clusters,  cluster matter introduces an
asymmetrical matter component superposing on cluster galaxies
as lenses. For simplicity, 
Turner et al (1984) used a uniform matter sheet as the model of 
a cluster. They inserted this additional mass density into the lensing
equation of a galaxy in the study of the multiple images of quasars.
This should be a good approximation if the ``lensing scale'' around 
the galaxy is much smaller than the cluster size. 
We essentially follow their methodology by approximating cluster
matter contribution as a uniform matter sheet superposed on galaxies.
Nevertheless, we  utilize a ``weighted'' mean cluster surface
mass density by considering the  galaxy distribution 
in clusters. To this end,
we assume an ISC profile with an one-dimensional velocity dispersion
$\sigma_{c}$ and a core radius $r_{cc}$ 
for the dark matter distribution of a cluster
\begin{equation}
\Sigma(\zeta)=\frac{\sigma_c^2}{2G}\frac{1}{\sqrt{\zeta^2+r_{cc}^2}}.
\end{equation}
Both the dynamical analysis of the X-ray observations and the study of
gravitational arclike images have shown that $r_{cc}$ should be 
much smaller than the core radius of the cluster luminous matter
(X-ray gas and galaxies) (e.g. Wu \& Hammer 1993; Durret et al.. 1994).
We will take $r_{cc}=0.1$ Mpc in our computations.  A numerical computation
shows that our final results remain almost unchanged if $r_{cc}$ varies from
0.05 to 0.25 Mpc.  For a cluster galaxy (SIS) at a radius  $\zeta$ 
from the cluster center, the magnification becomes then
\begin{equation}
\mu=\frac{\theta}{\theta-\theta_{cE}}
    \frac{1}{(1-\frac{\Sigma(\zeta)}{\Sigma_{crit}})^2},
\end{equation}
where $\theta_{cE}$ is the Einstein radius and is related to 
$\theta_E$ by SIS through 
$\theta_{cE}=\theta_E/(1-\Sigma(\zeta)/\Sigma_{crit})$.

An important parameter appeared in the above equation
is the critical surface mass density
\begin{equation}
\Sigma_{crit}=\frac{c^2D_s}{4\pi G D_d D_{ds}}.
\end{equation}
It is apparent from Eq.(15) that
only those clusters of galaxies whose surface mass densities 
are close to $\Sigma_{crit}$ can produce a significant effect.
While it is easy to show that
$\Sigma_{crit}$ is smaller in a $\lambda_0$ dominated universe 
than in an $\Omega_0$ dominated
one,  clusters of galaxies can act as more efficient lenses 
if the cosmological constant is nonzero.

We consider two kinds of models for the distribution of galaxies
along cluster radius: ISC and King models, which correspond to
the following variations of galaxy surface number density 
$\kappa_g(\zeta)$ with cluster radius:
\begin{equation}
\kappa_g(\zeta)\propto \left\{
\begin{array}{ll}
(\zeta^2+r_{cg}^2)^{-1/2}, & {\rm ISC};\\
(\zeta^2+r_{cg}^2)^{-1}, & {\rm KING},\\
\end{array} \right.
\end{equation}
where $r_{cg}$ is the core radius of the galaxy number density profile
which has been observationally determined  to be
$r_{cg}\approx0.25$ Mpc (Bahcall 1977).  Utilizing the same galaxy luminosity
distribution and composition as those in the above section, 
we give the quasar enhancement factor by all the galaxies in clusters  
with velocity dispersion of $\sigma_{c}$: 
\begin{equation}
\langle q\rangle = \frac{
\int_0^{z_s}4\pi D_d^2(1+z_d)^3dr_{prop,z_d}
\int_0^{R_g}\langle qn_g\rangle \kappa_g(\zeta)2\pi\zeta d\zeta}
{\int_0^{z_s}4\pi D_d^2(1+z_d)^3dr_{prop,z_d}
\int_0^{R_g}\langle  n_g\rangle \kappa_g(\zeta)2\pi\zeta d\zeta},
\end{equation} 
where $R_g$ denotes the cluster radius.

To include the contributions from different
clusters of galaxies.  we adopt
the cluster mass function established by Bahcall \& Cen (1993)
\begin{equation}
n_c(>M_c)=\phi^*_c(M_c/M_c^*)^{-1}\exp(-M_c/M_c^*),
\end{equation}
where $\phi^*_c$ is the normalization and 
$M_c^*=3.6\times10^{14}M_{\odot}$. $M_c$ refers to the cluster
mass within $R_g=3$ Mpc radius sphere of the cluster center and
$M_c=2\sigma_c^2R_g/G$ in SIS model. 
Note that we have used the same $R_g$ for the truncated radius 
of galaxy distribution as in Eq.(18).
The expected quasar enhancement factor around
cluster galaxies can be finally written as
\begin{equation}
\langle\langle q\rangle\rangle=
\frac{\int_{M_{c,min}}^{\infty}\langle q\rangle (dn_c/dM_c)dM_c}
{n_c(>M_{c,min})},
\end{equation}
in which $M_{c,min}$ is the low mass limit in the cluster mass function
and is taken to be $0.1M_c^*$ below.

\section{Numerical results}

In order to test the influence of different factors
on the prediction of the enhancement factor, we need to construct our 
``null-hypothesis''.   We use SIS as the mass density profile for
a galaxy which is described by the observed velocity dispersion of
stellar population (no bias). Furthermore, galaxies are assumed to be 
isolated objects in an $\Omega_0=1$ universe 
and non-evolved with cosmic epoch.  
We will calculate the expected enhancement factor by altering
one parameter each time in order to clearly demonstrate its effect on
$\langle q\rangle$.

1. Galactic morphologies. 
We first examine how the morphological composition of
galaxies affects the estimate of quasar enhancement factor. 
As is well known, E/S0 galaxies are more massive and hence more
efficient lenses than spirals. Thus, 
a higher quasar enhancement factor is provided by E/S0 galaxies.
In Table 2 We have given the theoretically expected 
enhancement factors by three types of galaxies separately
for each measurement.  The fraction of E, S0 and S field galaxies
remains roughly constant down to a relatively faint magnitude.  
For instance, even in the deep galaxy surveys to $B\sim22$ 
(e.g. Broadhurst et al. 1988) which is
comparable to the limiting magnitudes in the searches 
for the quasar-galaxy associations, one has found a similar
galaxy composition to what we have adopted in our null-hypothesis:
E:S0:S=12:19:69. However, this composition may vary in clusters
which contain more E/S0 galaxies than spirals.

2. Cosmic evolution.  
A question relevant to the morphological composition of
galaxies is the galaxy evolutionary effect. 
The merging model, i.e., our evolution model II motivated by  
a considerably large population of faint blue galaxies,
suggests more galaxies at high redshifts as
lenses [Eq.(10)], and almost all of the ellipticals may be the result of 
galaxy merging at $z\sim1$.  This indeed changes the morphological
components. While the velocity dispersion of galaxies was smaller in the past
according to the prediction of galaxy mergers, galaxies would
appear to be less efficient lenses as compared with our null hypothesis.
Therefore, the evolutionary model in which galaxies merge at recent
look-back time (Broadhurst at al. 1992) yields a relatively small 
quasar enhancement factor. On the other hands, the empirical 
form of the luminosity-dependent evolution (model I) predicts 
an increase of $\langle q\rangle$ because of the too many massive 
galaxies at high redshifts. Again,  $\langle q\rangle$ could be
significantly overestimated by this unphysical model.

3. Core radius. 
We now turn to the uncertainty in modeling of galaxy matter distributions. 
Instead of the unphysical condition of an infinite matter density in SIS
at the center of a galaxy, ISC is often invoked. The core radius
$r_c$ varies from galaxy to galaxy, depending on the galaxy
luminosity or velocity dispersion. Turner (1991) and Kochanek (1996)
adopted a relation $r_c\sim L^{\delta}$ where $\delta$ is determined
experientially. Here we utilize a mean core radius of $r_c=2$ kpc 
for all the galaxies. Our numerical results (Table 2) indicate that
the introduction of a definite core radius of $2$ kpc slightly reduces 
the value of $\langle q\rangle$.

4. Velocity bias.
Velocity dispersion is a critical parameter in the determination of
galaxy masses. It has been argued that
the observed velocity dispersion of stars in galaxies may not 
be representative of the dark matter behavior, i.e., the dark matter
may have a larger velocity dispersion $\sigma_{DM}$ than that observed, 
and there is a velocity biasing parameter $b$ between the dark matter and 
the stellar objects. 
This arises because the stellar population often follows a density profile
of $r^{-3}$ while the dark halo exhibits a form of $r^{-2}$ such as 
SIS and ISC. It has remained unclear whether the bias
parameter $b$ should be taken into account in the study of gravitational
lensing (Kochanek 1993; 1994). We adopt a value of $b=\sqrt{1.5}$
for E/S0 galaxies (Turner et al. 1984; Fukugita \& Turner 1991) to
illustrate how our prediction of $\langle q\rangle$ is affected
by this uncertainty. As shown in Table 2, the expected 
$\langle q\rangle$ with and without the bias parameter are indeed different:
the correction of velocity dispersion by a factor of $b$ would evidently 
increase the value of $\langle q\rangle$.

5. Clusters of galaxies.
The fraction of all galaxies that belong in clusters or in fields
is quite uncertain, which prevents us from quantitatively setting a plausible 
mixture of the cluster galaxies and the field galaxies. 
If the fraction of galaxies in rich clusters is only $\sim5\%$ 
(Bahcall 1996), then the cluster contribution to the quasar-galaxy 
associations may become trivial. 
Here we discuss an extreme case in which all the 
galaxies are bounded in clusters. 
Namely, we evaluate the maximum contribution of the galaxy environmental 
matter from clusters to the quasar-galaxy associations. 
Essentially, clusters provide an additional matter 
component to the galactic lenses and raise
the quasar enhancement factor. However, 
as the mean cluster surface mass density is considerably
smaller than the critical value $\Sigma_{crit}$ for most of the clusters 
when $\zeta>r_{cc}$, 
our numerical computations indicate that the environmental matter of cluster
galaxies produce little effect on $\langle q\rangle$.

6. Obstruction. 
The obstruction effect by the luminous disk of foreground galaxies 
turns to be negligible:  The largest effect leads to a decrease of
$\langle q\rangle$ only by $\sim0.1$ for Magain's observation, whereas 
others are nearly unaffected by obstruction since their
searching distances are well beyond the galactic luminous disks.

7. Cosmological constant. 
It appears that a cosmological constant dominant universe of 
$\lambda_0=0.8$ ($\Omega_0=0.2$) 
does not provide a significant difference in the prediction of
$\langle q\rangle$ from a matter dominant one 
($\lambda_0=0$ and $\Omega_0=1$), and the increased value of 
$\langle q\rangle$ by a non-zero $\lambda_0$ is rather small (see Table 2).
This is consistent with the early analysis by Fukugita et al (1992). 
The simple reasons are as follows: $\lambda_0$ enters into 
$\langle q\rangle$ through the Einstein radius 
$\theta_E=4\pi(\sigma_{DM}/c)^2(D_{ds}/D_s)$ 
and the volume element $dV=4\pi D_d^2dr_{prop,z_d}$. While 
$D_{ds}/D_s\approx1$ for the measurement of quasar-galaxy associations,
$\theta_E$ is roughly independent of the cosmological models.
On the other hand, the contribution of $\lambda_0$ is depressed
when $\int\langle qn_g\rangle dV/\int\langle n_g\rangle dV$ [Eq.(11)]
is employed.

8. Quasar number counts. We have not utilized the widely adopted
quasar number count in literature, namely, the quasar number-magnitude
relation found by Boyle, Shanks and Peterson (1988). Instead, we have 
adopted a combination of the quasar number counts by numerous observations. 
As a consequence, the slope of $d\log N_q/dB$ is somewhat increased at
both bright and faint ends of quasar magnitude. We have compared  
the resulted $\langle q\rangle$ from the Boyle et al. (1988)  counts and 
Eq.(2), and found that the difference is minor. Alternatively,
all the results in Table 2 correspond to the quasar redshift limit of 
$z_s<2.2$. Noticing that some measurements in Table 1 (e.g. Magain and 
Van Drom) may contain a large fraction of high redshift ($z_s>2.2$)
quasars, we have also tested the quasar number-magnitude relation
$N_q(<B)$ for $2.2<z_s<3.0$ [see Eq.(2)]. This leads to a decrease of
the prediction of $\langle q\rangle$  because of
the flattening of the $N_q(<B)$.

\section{Discussion and conclusions}

Unlike the previous statistical studies on the quasar-galaxy associations,
we have not employed the magnification probability
function (e.g. Schneider 1989) in the present paper.
This reduces the complexity of
computations and avoids the arbitrary choice of a low
magnification limit in the convolution of the magnification
probability function with the quasar number count or 
luminosity function. While the quasar enhancement factor around
a single galaxy is known,  we have statistically obtained the 
expected quasar enhancement factor $\langle q\rangle$ around foreground
galaxies by averaging $q$ over galactic morphologies, luminosities and 
redshifts. Moreover, we have  included the contributions of
the environmental matter surrounding galaxies from their host
clusters. Other effects such as the possible bias between 
the velocity dispersion of the stellar population and of the dark matter,
the galaxy evolutionary effect 
and a non-zero cosmological constant have also been 
considered. As a whole, we have made an extensive theoretical study 
and have presented an updated estimate of the amplitude of the
quasar-galaxy associations in terms of our best knowledge today. 
Table 2 summarizes the measured and expected values of the   
quasar enhancement factors for nine observations.

Overall, as compared with observations, 
galaxies alone provide a relatively small 
quasar enhancement factor $\langle q\rangle$.
Among various factors and uncertainties we have studied,  
the following three parameters may produce the most significant 
effect on the estimates of $\langle q\rangle$:
the existence of a velocity bias between the stellar objects and the dark
matter, the non-zero cosmological constant and the cluster
matter contributions. However, their induced
variations in $\langle q\rangle$ are still minor, which 
cannot increase $\langle q\rangle$ to the values
that agree with  all the observations.   
It appears that the combined result of some affects 
may marginally account for the observed  
quasar enhancement factor, if a large observational uncertainty
is presumed (e.g. Webster's measurement). 
The results from a combination of 
a velocity bias $b$  and a non-zero cosmological constant of $\lambda_0=0.8$
has been shown in Table 2.

The gravitational lensing mechanism  may be able 
to well reproduce the so-called ``null'' 
or negative results of the quasar-galaxy associations,
but has less power for the explanation of the 
large $\langle q\rangle$ events. 
If the reported quasar-galaxy associations are not due to
statistical variations or suffer from other selection effects, 
this might imply that a steeper quasar number count is required. 
It is not impossible that the slope of the quasar number-magnitude 
relation $N_q(<B)$ can be as large as $\sim0.4$ at the faint 
limiting magnitude
because there is still sufficient uncertainty in the present 
quasar number counts.  Alternatively, the question remains open whether the
observed quasar counts have already been contaminated by lensing.
Recall that the gravitational lensing explanation shows a 
similar inefficiency when applied for the quasar-cluster associations
(Wu \& Fang 1996).

Yet, the observational selection effects in the measurement of the 
quasar-galaxy associations are quite complex and error bars have 
not been given for some of the observations especially for those  
large enhancement factors. This makes the comparison of the
theoretical expectation and the observation very difficult. Because
the detections of the ``null'' result at the faint quasar limiting magnitude
and the positive result at the bright one in the measurements of
quasar-galaxy associations are basically consistent 
with the scenario of gravitational lensing and because it is 
lack of convincing evidences to support other explanations
such as the physical associations, 
we still believe that the quasar-galaxy associations are relevant 
to gravitational lensing. However, our detailed examinations of
the lensing models indicate that either current measurements 
are unreliable or we need to modify at least one of the basic
hypotheses in the lensing explanation for the quasar-galaxy associations.


\acknowledgments

We thank B. Qin for careful reading the manuscript and  
an anonymous referee for valuable comments and suggestions.  
WXP was supported by the National Science Foundation
of China and a World-Laboratory fellowship.


\begin{center}
\begin{table}
\caption{Quasar-galaxy associations: observations$^*$
\label{table-1}}
\scriptsize
\bigskip
\begin{tabular}{ccccccc}
\tableline
\tableline
labels & authors & QSO No.   & selections   &
$\theta$ range ($^{\prime\prime}$) & galaxies ($R$)  &
observed  $q_{obs}$\\
\tableline
C & Crampton &
101 & $V<18.5$, $z>1.5$ & 0 -- 6  & $\sim23$   & $1.4\pm0.5$ \nl
K & Kedziora-Chudczer &
181 & $V<18.5$, $z>0.65$ & 6 -- 90 & $\sim21.5$ & $\sim1$     \nl
M & Magain   &
153 & $\langle V\rangle=17.4$, $\langle z\rangle=2.3$ & 0 -- 3 &
$\sim21$ & $\sim2.8$     \nl
T & Thomas &
 64 & $V<18.5$, $1<z<2.5$ & 0 -- 10 & $\sim22$ & $1.7\pm0.4$   \nl
V & Van Drom &
136 & $\langle V\rangle=17.4$, $\langle z\rangle=2.3$ & 3 -- 13.7 &
$\sim 23$ & $\sim1.46$     \nl
W & Webster &
 68 & $V<18$, $0.7<z<2.3$ & 3 -- 10 & $\sim22$ & $\sim2$     \nl
Y1&       &
    &                     & 2 -- 6  &          & $1.0\pm0.3$     \nl
Y2 & Yee   &
94  & $V<19$, $z>1.5$     & 2 -- 10 & $\sim22.5$ & $1.0\pm0.2$   \nl
Y3 &    &
    &                     & 2 -- 15 &          & $0.9\pm0.1$    \nl
\tableline
\end{tabular}
\tablenotetext{*}{Data are taken from Narayan (1992) and Wu (1996).} 
\end{table}
\end{center}

\clearpage


\begin{table}
\caption{Quasar-galaxy associations: expectations
\label{table-2}}
\scriptsize
\bigskip
\begin{center}
\begin{tabular}{cccccccccc}
\tableline
\tableline
labels & C & K & M & T & V & W & Y1 & Y2 & Y3\\
\tableline
$q_{obs}$ & $1.4\pm0.5$ &  $\sim1$ & $\sim2.8$ & $1.7\pm0.4$ & $\sim1.46$ & $\sim2$ &  $1.0\pm0.3$ &  $1.0\pm0.2$ &  $0.9\pm0.1$ \\
(1) & $ 1.20^{ +0.04}_{ -0.04}$ & $ 1.02^{ +0.00}_{ -0.01}$ & $ 1.97^{ +0.24}_{ -0.25}$ & $ 1.15^{ +0.03}_{ -0.03}$ & $ 1.08^{ +0.02}_{ -0.02}$ & $ 1.12^{ +0.04}_{ -0.03}$ & $ 1.07^{ +0.01}_{ -0.01}$ & $ 1.05^{ +0.01}_{ -0.01}$ & $ 1.03^{ +0.01}_{ -0.00}$

\\ 
(2) & $ 1.26^{ +0.06}_{ -0.05}$ & $ 1.02^{ +0.01}_{ -0.00}$ & $ 2.20^{ +0.31}_{ -0.32}$ & $ 1.18^{ +0.05}_{ -0.04}$ & $ 1.10^{ +0.04}_{ -0.02}$ & $ 1.16^{ +0.04}_{ -0.05}$ & $ 1.08^{ +0.02}_{ -0.02}$ & $ 1.06^{ +0.01}_{ -0.02}$ & $ 1.04^{ +0.01}_{ -0.01}$

\\
(3)& $ 1.22^{ +0.05}_{ -0.06}$ & $ 1.02^{ +0.00}_{ -0.01}$ & $ 1.98^{ +0.29}_{ -0.18}$ & $ 1.15^{ +0.04}_{ -0.04}$ & $ 1.09^{ +0.02}_{ -0.03}$ & $ 1.13^{ +0.04}_{ -0.04}$ & $ 1.07^{ +0.02}_{ -0.02}$ & $ 1.05^{ +0.01}_{ -0.01}$ & $ 1.04^{ +0.01}_{ -0.00}$
\\
(4) & $ 1.16^{ +0.03}_{ -0.03}$ & $ 1.01^{ +0.01}_{ -0.00}$ & $ 1.74^{ +0.13}_{ -0.17}$ & $ 1.12^{ +0.02}_{ -0.02}$ & $ 1.06^{ +0.01}_{ -0.01}$ & $ 1.10^{ +0.01}_{ -0.02}$ & $ 1.06^{ +0.00}_{ -0.01}$ & $ 1.04^{ +0.00}_{ -0.01}$ & $ 1.03^{ +0.00}_{ -0.01}$

\\
(5) & $ 1.21^{ +0.03}_{ -0.04}$ & $ 1.02^{ +0.00}_{ -0.01}$ & $ 2.01^{ +0.25}_{ -0.26}$ & $ 1.16^{ +0.02}_{ -0.04}$ & $ 1.08^{ +0.02}_{ -0.01}$ & $ 1.13^{ +0.03}_{ -0.03}$ & $ 1.07^{ +0.01}_{ -0.01}$ & $ 1.05^{ +0.01}_{ -0.01}$ & $ 1.03^{ +0.01}_{ -0.00}$

\\
(6) & $ 1.12^{ +0.02}_{ -0.03}$ & $ 1.01^{ +0.00}_{ -0.00}$ & $ 1.60^{ +0.16}_{ -0.17}$ & $ 1.09^{ +0.01}_{ -0.03}$ & $ 1.04^{ +0.01}_{ -0.01}$ & $ 1.07^{ +0.02}_{ -0.02}$ & $ 1.04^{ +0.01}_{ -0.01}$ & $ 1.03^{ +0.01}_{ -0.01}$ & $ 1.04^{ +0.01}_{ -0.01}$

\\
(7) & $ 1.19^{ +0.04}_{ -0.05}$ & $ 1.02^{ +0.00}_{ -0.01}$ & $ 1.71^{ +0.17}_{ -0.19}$ & $ 1.14^{ +0.03}_{ -0.04}$ & $ 1.08^{ +0.02}_{ -0.02}$ & $ 1.12^{ +0.03}_{ -0.03}$ & $ 1.07^{ +0.01}_{ -0.01}$ & $ 1.05^{ +0.01}_{ -0.01}$ & $ 1.03^{ +0.01}_{ -0.00}$

\\
(8) & $ 1.26^{ +0.05}_{ -0.05}$ & $ 1.02^{ +0.01}_{ -0.00}$ & $ 2.35^{ +0.20}_{ -0.31}$ & $ 1.20^{ +0.04}_{ -0.04}$ & $ 1.12^{ +0.03}_{ -0.04}$ & $ 1.18^{ +0.05}_{ -0.05}$ & $ 1.08^{ +0.01}_{ -0.02}$ & $ 1.06^{ +0.01}_{ -0.01}$ & $ 1.04^{ +0.01}_{ -0.00}$

\\
(9) & $ 1.23^{ +0.04}_{ -0.05}$ & $ 1.04^{ +0.02}_{ -0.03}$ & $ 2.04^{ +0.26}_{ -0.28}$ & $ 1.18^{ +0.06}_{ -0.05}$ & $ 1.12^{ +0.04}_{ -0.05}$ & $ 1.16^{ +0.06}_{ -0.05}$ & $ 1.07^{ +0.02}_{ -0.01}$ & $ 1.06^{ +0.02}_{ -0.01}$ & $ 1.04^{ +0.01}_{ -0.00}$

\\

(10) & $ 1.19^{ +0.04}_{ -0.04}$ & $ 1.02^{ +0.00}_{ -0.01}$ & $ 1.80^{ +0.18}_{ -0.22}$ & $ 1.14^{ +0.03}_{ -0.03}$ & $ 1.08^{ +0.02}_{ -0.02}$ & $ 1.12^{ +0.03}_{ -0.03}$ & $ 1.06^{ +0.01}_{ -0.01}$ & $ 1.04^{ +0.01}_{ -0.00}$ & $ 1.03^{ +0.01}_{ -0.00}

$\\
(11) & $ 1.26^{ +0.04}_{ -0.06}$ & $ 1.02^{ +0.01}_{ -0.00}$ & $ 2.20^{ +0.24}_{ -0.31}$ & $ 1.19^{ +0.04}_{ -0.04}$ & $ 1.11^{ +0.02}_{ -0.03}$ & $ 1.16^{ +0.04}_{ -0.04}$ & $ 1.08^{ +0.02}_{ -0.01}$ & $ 1.06^{ +0.01}_{ -0.01}$ & $ 1.04^{ +0.01}_{ -0.01}

$\\
(12) & $1.33^{+0.04}_{-0.07}$ & $1.03^{+0.01}_{-0.01}$ & $2.57^{+0.10}_{-0.29}$
& $1.25^{+0.04}_{-0.05}$ & $1.16^{+0.04}_{-0.05}$ & $1.25^{+0.07}_{-0.07}$ &
  $1.10^{+0.01}_{-0.02}$ & $1.07^{+0.01}_{-0.01}$ & $1.05^{+0.01}_{-0.01}$ \\
\tableline
\end{tabular}
\end{center}

\footnotesize
\tablenotetext{1}{Null-hypothesis: E:S0:S=12:19:69, SIS model, no velocity
                  bias, no-evolution and $\lambda_0=0$. The error bars are
                  the combined result of the uncertainties in the quasar number
                  counts $N_q(S)$, the luminosity function $\phi(L,z)$ and
                  the characteristic velocity dispersion $\sigma_*$.}  
\tablenotetext{2}{E galaxies as lenses only.}
\tablenotetext{3}{S0 galaxies as lenses only.}
\tablenotetext{4}{S galaxies as lenses only.}
\tablenotetext{5}{Evolution model I: the luminosity-dependent evolution.}
\tablenotetext{6}{Evolution model II: the merging model.}
\tablenotetext{7}{ISC with a core radius of $r_c=2$ kpc.}
\tablenotetext{8}{With a biasing parameter $b=(3/2)^{1/2}$ for velocity
                  dispersion of E/S0 galaxies.}
\tablenotetext{9}{Environmental effect from cluster matter contribution.}
\tablenotetext{10}{Obstruction effect by foreground galactic disks.}
\tablenotetext{11}{The universe with a non-zero cosmological constant of 
                   $\lambda_0 = 0.8$.}
\tablenotetext{12}{Combined result of a velocity bias $b$ and a non-zero
                   cosmological constant of $\lambda_0 = 0.8$.}  
\end{table}

\clearpage


\end{document}